\documentclass[aps,twocolumn,showpacs,
superscriptaddress,groupedaddress,
floatfix,
twocolumn,aps,prd,amsmath,amssymb]{revtex4}
\usepackage{epsfig}
\usepackage{multirow}
\usepackage{graphicx}
\usepackage{graphics}
\usepackage{subfigure}
\usepackage{epstopdf}
\usepackage{float}
\usepackage{bm}
\usepackage{bbm}
\usepackage{braket}
\usepackage{lineno}
\usepackage{slashed}
\usepackage{hyperref}
\hypersetup{colorlinks=true,    linkcolor=red,filecolor=black,
citecolor=blue}

\begin{document}

\title{Dynamical Mass Generation in Pseudo Quantum Electrodynamics with Gross-Neveu Interaction at finite temperature}

\author{Luis Fern\'andez}
\email{luis.aguilar@icen.ufpa.br} \affiliation{Faculdade de F\'\i sica, Universidade Federal do Par\'a, 66075-110, Bel\'em, PA,  Brazil}

\author{Reginaldo O. Corr\^ea Jr.}
\email{reginaldojunior@uepa.br} \affiliation{Centro de Ci\^encias e Planet\'ario Sebasti\~ao Sodr\'e da Gama, Universidade do Estado do Par\'a, Rodovia Augusto Montenegro km 03, 66640-000, Bel\'em, PA, Brazil}

\author{Van S\'ergio Alves}
\email{vansergi@ufpa.br}
\affiliation{Faculdade de F\'\i sica, Universidade Federal do Par\'a, 66075-110, Bel\'em, PA,  Brazil}

\author{Leandro O. Nascimento}
\email{lon@ufpa.br}
\affiliation{Faculdade de Ci\^encias Naturais, Universidade Federal do Par\'a, C.P. 68800-000, Breves, PA,  Brazil}

\author{Francisco Pe\~na}
\email{francisco.pena@ufrontera.cl} 
\affiliation{Departamento de Ciencias F\'\i sicas, Facultad de Ingenier\'\i a y Ciencias, Universidad de La Frontera, Avda. Francisco Salazar 01145, Casilla 54-D, Temuco, Chile}

\date{\today}

\begin{abstract}
We study the dynamical mass generation in Pseudo Quantum Electrodynamics (PQED) coupled to the Gross-Neveu (GN) interaction, in (2+1) dimensions, at both zero and finite temperatures. We start with a gapless model and show that, under particular conditions, a dynamically generated mass emerges. In order to do so, we use a truncated Schwinger-Dyson equation, at the large-N approximation, in the imaginary-time formalism. In the instantaneous-exchange approximation (the static regime), we obtain two critical parameters, namely, the critical number of fermions $N_c(T)$ and the critical coupling constant $\alpha_c(T)$ as a function of temperature and of the cutoff $\Lambda$, which must be provided by experiments. In the dynamical regime, we find an analytical solution for the mass function $\Sigma(p,T)$ as well as a zero-external momentum solution for $p=0$. We compare our analytical results with numerical tests and a good agreement is found.
\end{abstract}

\pacs{}

\maketitle

\section{\textbf{Introduction}}

In the last decades, the interest in studying two-dimensional theories has been increased since the experimental realization of graphene \cite{grapexp,review} and other related materials, such as the transition metal dichalcogenide monolayers \cite{TMDs}. The main goal is to derive either new quantum phases of matter \cite{Haldane,Bernevig,QSHE,prx} or calculate renormalized parameters that change the electronic properties of these materials \cite{Voz1,foster,Gui,popovici,geim, PQEDGap}, opening possibilities for future technological applications (in particular spintronics \cite{spintronicsnature}, valleytronics \cite{valleytronics}, and electric-field tunning of energy bands \cite{WSe2gap,MoS2gap}). Before this context, in the realm of high-energy physics, several works discussed the possibility of dynamical mass generation for massless Dirac particles, yielding a phase transition to a new quantum state of matter in which the chiral symmetry is broken \cite{maris,robert,curtis,appelquist}. This is generated due to the electronic interactions in the plane and it may occur even at finite temperatures. Because electrons in these materials obey a Dirac-like equation, therefore, some authors have also considered the realization of dynamical mass generation in these two-dimensional systems \cite{Herbut,VLWJF,ChunXu,teber1}. 

The possibility of chiral symmetry breaking, i.e, dynamical mass generation in quantum electrodynamics (QED) has been discussed in several Refs. \cite{maris,bashir1,bashir2,bloch,robert,curtis,appelquist} for both (2+1)D and (3+1)D cases, just to cite a few. Furthermore, this symmetry-broken phase also has been shown to occur due to four-fermion interactions (such as Gross-Neveu and Thirring interactions) in Refs.~\cite{akram, gomes1, GNT=0, semmenoff}. In this case, the fermionic field exhibits a massive phase due to the spontaneous symmetry breaking, described by the so-called gap equation. This is usually calculated at the large-$N$ expansion, where $N$ is the number of fermion fields and the gap equation is calculated at order of $1/N$ \cite{akram,gomes1}. The dynamical mass generation is a typical effect of the nonperturbative regime, hence, it is common the application of the Schwinger-Dyson equations (SDEs) for calculating the critical parameters that describe the phase transition. 

The SDEs are an infinite set of coupled-integral equations, relating all of the interacting Green functions of the model \cite{curtis, robert, fayzullaev}. Fortunately, within some approximative scheme, it is possible to find a truncated set of equations for calculating the desired Green functions, in particular, the electron-self energy that provides the chiral symmetry breaking. For pseudo-quantum electrodynamic (PQED) \cite{marino,VLWJF}, also called reduced quantum electrodynamics \cite{teber0}, the dynamical mass generation has been studied both at finite temperatures \cite{PQEDT} and at zero temperature with the presence of the Gross-Neveu interaction \cite{GNT=0}. Nevertheless, the effect of the Gross-Neveu interaction in PQED at finite temperatures has not been considered until now.

In this work, we describe the dynamical mass generation in PQED coupled to a Gross-Neveu interaction at finite temperature. We use the Matsubara formalism in order to include the effect of the thermal bath into the Schwinger-Dyson equation for the electron. This is dependent on both the gauge-field propagator and the auxiliary-field propagator, obtained after we use a Hubbard-Stratonovich transformation into the four-fermion interaction. These two bosonic propagators are calculated in order of $1/N$, which is consistent with our assumption of strong-coupling limit. Thereafter, we use this result into the Schwinger-Dyson equation for the electron self-energy and calculate the full electron propagator in the dominant order of $1/N$ in the nonperturbative limit. From this result, we conclude that a mass function is dynamically generated whether the number of fermions is less than a temperature-dependent critical parameter $N_c(T)$. At large temperatures, we find that $N_c(T)\ll 1$, hence, the dynamically generated mass vanishes and the system is in the gapless phase.

This paper is organized as follow. In Sec. II. we show our model and perform the large-$N$ approximation. In Sec. III.  we write the truncated set of Schwinger-Dyson equations within the unquenched-rainbow approximation. In Sec. IV. we calculate the mass function in the static regime and obtain the critical parameters for the phase transition.  In Sec.V. we use the zero-mode approximation in the dynamical regime for calculating the mass function. In Sec.VI. we summarize and discuss our main results. We also include three appendices, where we give details about the angular integral, kernel expansion, and the numerical results for the mass function.

\section{\textbf{Pseudo-quantum electrodynamics with Gross-Neveu interaction}}

We consider $N$ fermion fields constrained to the plane, whose interaction is described by Pseudo-quantum electrodynamics (PQED) \cite{marino}. Furthermore, we assume that these particles  involve a contact interaction, given by the Gross-Neveu (GN) action  \cite{gross}. Therefore, in the Euclidean space-time, the action of the model reads     
\begin{equation}
\begin{split}
{\cal L}&=\frac{1}{2} \frac{F_{\mu \nu}F^{\mu\nu}}{\sqrt{-\Box}} +i \bar\psi_a\partial\!\!\!/ \psi_a + e \bar{\psi}_a\gamma^\mu \psi_a A_{\mu} + \\ &  -\frac{G_0}{2}(\bar\psi_a \psi_a)^2 - \frac{\xi}{2}A_{\mu}\frac{\partial^{\mu}\partial^{\nu}}{\sqrt{-\Box}}A_{\nu},
\label{action}
\end{split}
\end{equation}
where $F_{\mu \nu}$ is the field intensity tensor of $A_{\mu}$, which is the gauge field, $\xi$ is the gauge-fixing parameter, $\psi_a$ is a four-component Dirac field, with the flavor index $a = 1, \ldots, N$, the dimensionless coupling constant $e$ is the electric charge, $\gamma_{\mu}$ are the Dirac matrices in the four-rank representation, whose algebra is defined by $\left\lbrace \gamma_{\mu}, \gamma_{\nu} \right\rbrace = -2 \delta_{\mu \nu}$, the coupling constant $G_0$ describes the strength of the GN interaction and has unit of mass in the natural system of units $(\hbar=c =1)$. 

In the perturbative regime at zero temperature, the model in Eq.~(\ref{action}) for massive fermions has been used to describe the renormalization of the bandgap for WSe$_2$ and MoS$_2$ \cite{PQEDGap}. This result, nevertheless, requires a bare-mass term such as $m_0 \bar\psi_a \psi_a$ which is renormalized at one-loop and provides a beta function for the mass in terms of the RG scale. From this result, one concludes that the renormalized mass is dependent on the electronic density and a good agreement with experimental data has been found in Ref.~\cite{PQEDGap}

In the nonpertubative limit, the mass term is generated due to interactions even if we start with $m_0\rightarrow 0$ \cite{GNT=0}. Here, we generalize this result by including a thermal bath of temperature $T$. 
 
From Eq.~(\ref{action}), we find the gauge-field propagator, namely,
\begin{equation}
\Delta_{0,\,\, \mu\nu}(p)=\frac{1}{2\sqrt{p^2}} \left[\delta_{\mu\nu} - \left(1-\frac{1}{\xi}\right) \frac{p_{\mu}p_{\nu}}{p^2} \right]  \label{photonbare}
\end{equation}
and the fermion propagator
\begin{equation}
S_{0,\,F}(p)=-\frac{1}{\gamma^\mu p_\mu}.
\label{fermionfree}
\end{equation}

Before we discuss the vertex interactions, let us apply the Hubbard-Stratonovich transformation, which converts the four-fermion interaction into a Yukawa-type interaction by including an auxiliary field $\varphi$ and the coupling constant $g=G_0N$.  In this case, we replace the four-fermion interaction by the following scheme   
  \begin{equation}
{\cal L}_{\rm{GN}}\rightarrow {\cal L}_{\rm{GN}}+\frac{1}{2g}\left(\varphi-\frac{g}{\sqrt{N}}\bar\psi_a\psi_a\right)^2, \label{transf}
\end{equation}
where ${\cal L}_{\rm{GN}}=-G_0 (\bar\psi_a \psi_a)^2/2$. Therefore, after that we find  
\begin{equation}
{\cal L}_{\rm{GN}}= -\frac{\varphi}{\sqrt{N}} \bar\psi_a\psi_a+\frac{\varphi^2}{2g}. 
\label{action2}
\end{equation}
Equation \eqref{action2} must be supplemented by the motion equation of $\varphi$, given by
\begin{equation}
\varphi=\frac{g}{\sqrt{N}}\bar\psi_a\psi_a,
\end{equation}
which proves that the transformation does not change the dynamics of the model at classical level. Furthermore, the bare auxiliary-field propagator is
\begin{equation}
\Delta_{0, \,\varphi}=\frac{1}{1/g},
\end{equation}
and Eq. \eqref{action} reads
\begin{equation}
\begin{split}
{\cal L}&= \frac{1}{2} \frac{F_{\mu \nu}F^{\mu\nu}}{\sqrt{-\Box}} + i\bar\psi_a\partial\!\!\!/ \psi_a+e\bar{\psi}_a \gamma^{\mu}\psi_a A_{\mu} \\
&-\frac{\varphi}{\sqrt{N}} \bar\psi_a\psi_a+\frac{\varphi^2}{2g}- \frac{\xi}{2}A_{\mu}\frac{\partial^{\mu}\partial^{\nu}}{\sqrt{-\Box}}A_{\nu}.
\label{actionend}
\end{split}
\end{equation}
The Yukawa-type vertex function is easily obtained from Eq. \eqref{actionend} and is given by $-1/\sqrt{N}$. On the other hand, for summing the self-energies in the large-$N$ expansion for PQED, we shall replace $e^2\rightarrow \lambda/N$, where $\lambda$ is taken fixed at large $N$. This allow us to sum over all of the diagrammatic contributions in order of $1/N$, which is an infinite sum, unlike the standard perturbation in $e$. Therefore, the PQED vertex reads $\sqrt{\lambda/N} \gamma_\mu$, describing the electromagnetic interaction.
 
\section{ \textbf{Truncated Schwinger-Dyson equation at finite temperatures}}

In this section we present the Schwinger-Dyson equations, obtained from Eq. \eqref{actionend}, that describes the quantum corrections for the two-point functions. In principle, this is a very complicated set of  integral equations for all of the full Green functions of the model. From now on, we assume the ladder approximation, also called rainbow approach \cite{rainbow}, which consists of neglecting quantum corrections to the vertex functions. It is worth to note that this may be corrected by the Ball-Chiu vertex \cite{ball} when one wishes to preserve the Ward-Takahashi identity.

Because we are interested in dynamical mass generation for the electrons, we need to find a closed solution for both gauge-field and auxiliary-field propagators. These shall be given by the large-$N$ approximation, thus the bosonic-field propagators are calculated in the unquenched approximation \cite{kondo}.

\subsection{ \textbf{Auxiliary field}}

\begin{figure}[H]
\centering
\includegraphics[scale=0.8]{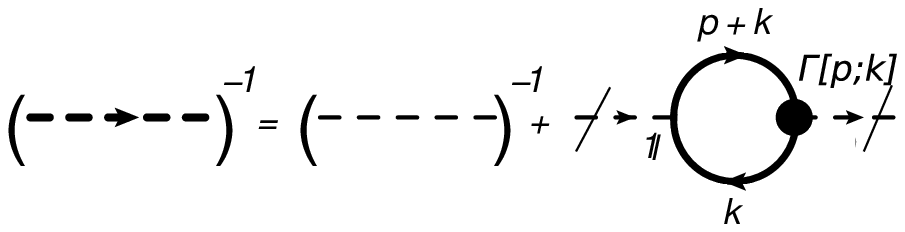}
\caption{The Schwinger-Dyson equation for the auxiliary field. The left-hand side is the inverse of the full propagator of the auxiliary field, the first term in the right-hand side corresponds to the bare auxiliary-field propagator, and the second term $\Pi(p)$ is the quantum correction. The function $\Gamma[p,k]\rightarrow \mathbbm{1}$ corresponds to the approximated vertex.} \label{Fig1}
\end{figure}

In Fig. \ref{Fig1}, we show the diagrammatic representation of the Schwinger-Dyson equation for the auxiliary field.
Its analytical expression is given by \cite{GNT=0}
\begin{equation}
\Delta^{-1}_\varphi(p)=\Delta^{-1}_{0, \,\varphi}-\Pi(p),
\end{equation}
where $\Delta^{-1}_\varphi$ is the full propagator of $\varphi$ and $\Pi(p)$ is given by the fermionic loop, hence
\begin{equation}
\Pi(p)=-\frac{1}{N}\textrm{Tr} \int\frac{d^3k}{(2\pi)^3}\mathbbm{1} S_F(p-k)\Gamma(p,k) S_F(k),
\end{equation} 
where $S_F(p)$ is the full fermion propagator. In the lowest order of $1/N$, we find that $\Pi(p)=-\sqrt{p^2} g_0$, with $g_0=1/4$ \cite{appelquist}. Therefore, the full propagator of the auxiliary field is given by
\begin{equation}
\Delta_\varphi(p)=\frac{1}{g_0}\frac{1}{(gg_0)^{-1}+\sqrt{p^2}}. \label{aux}
\end{equation}

\subsection{ \textbf{Gauge field}}

\begin{figure}[H]
\centering
\includegraphics[scale=0.8]{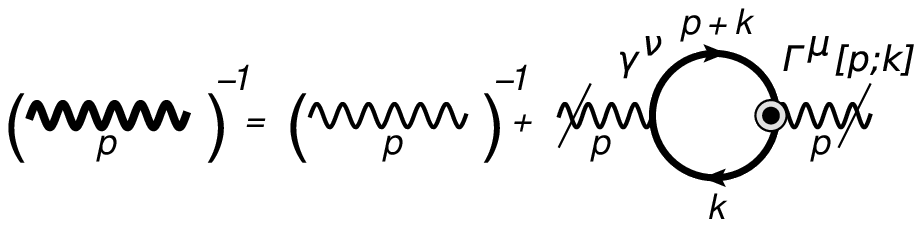}
\caption{The Schwinger-Dyson equation for the gauge-field propagator. The left-hand side is the inverse of the full propagator of the gauge field, the first in the right-hand side corresponds to the bare gauge-field propagator, and the second term $\Pi_{\mu\nu}(p)$ is the exact vacuum polarization tensor. The function $\Gamma^{\mu}[p,k]\rightarrow \gamma^{\mu}$ corresponds to the approximated vertex.} \label{Fig2}
\end{figure}

In Fig.~(\ref{Fig2}), we show the diagrammatic representation of the Schwinger-Dyson equation for the gauge-field propagator \cite{GNT=0}.
Its analytical expression is given by
\begin{equation}
\Delta^{-1}_{\mu\nu}(p)=\Delta^{-1}_{0,\,\mu\nu}(p)-\Pi_{\mu\nu}(p),
\label{SDphoton}
\end{equation}
where $\Delta^{-1}_{\mu\nu}$ is the full propagator and $\Pi_{\mu \nu}$ is the polarization tensor due to the electromagnetic interaction. This is given by 
\begin{equation}
\Pi_{\mu \nu}(p) = \!- \frac{\lambda}{N} \textrm{Tr}\!\! \int\!\! \frac{d^3 k}{(2 \pi)^3} \gamma_{\mu} S_F(p+k)\Gamma_{\nu}(p,k)S_F(k).
\label{polgauge}
\end{equation}
Similarly to the previous case, in the lowest order of $1/N$, we find $\Pi_{\mu\nu}= \lambda \sqrt{p^2}/8 P_{\mu \nu}$, where $P_{\mu \nu}= \delta_{\mu \nu}-p_{\mu}p_{\nu}/p^2$. Therefore, the full propagator reads  
\begin{equation}
\Delta_{\mu\nu}=\Delta_{0, \,\mu\alpha}[\delta^{\alpha}_\nu-\Pi^{\alpha\beta}\Delta_{0,\,\beta\nu}]^{-1}. \label{solphoton}
\end{equation}
Using Eqs. \eqref{photonbare} and \eqref{polgauge} in Eq. \eqref{solphoton}, we find
\begin{equation}
\Delta_{\mu\nu}(p)=\frac{P_{\mu\nu}}{\sqrt{p^2}\left(2+\frac{\lambda}{8}\right)}
\label{gaugefull}
\end{equation}
in the Landau gauge, i.e, with $\xi=\infty$.

\subsection{ \textbf{Matter field}}
\begin{figure}[H]
\centering
\includegraphics[scale=0.8]{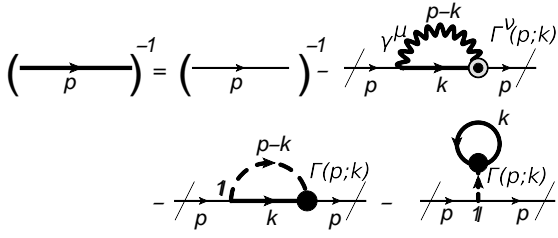}
\caption{The Schwinger-Dyson equation for the electron. The left-hand side is the inverse of the full propagator of the fermion field, the first term in the right-hand side corresponds to the bare Dirac propagator, and the other terms in the right-hand side are the electron self-energy $\Xi(p)$.} \label{Fig3}
\end{figure}

In Fig.~(\ref{Fig3}), we show the diagrammatic representation of the Schwinger-Dyson equation for the fermions field \cite{GNT=0}.
Its analytical expression is given by
\begin{equation}
S_F^{-1}(p)=S_{0,\,F}^{-1}(p)-\Xi(p),
\label{sdefermion}
\end{equation}
where $S_F$ is the full fermion propagator and $\Xi$ is the electron self-energy, given by 
\begin{equation}
\Xi(p) = \Xi^{\lambda}(p) + \Xi^{g}(p),
\end{equation}
where $\Xi^{\lambda}$ is the PQED contribution and $\Xi^{g}$ is the GN contribution, namely,
\begin{equation} \label{Xi1}
\Xi^{\lambda}(p)= \frac{\lambda}{N} \int \frac{d^3 k}{(2 \pi)^3} \gamma^{\mu}S_F(k)\Gamma^{\nu}(k;p)\Delta_{\mu \nu}(p-k),
\end{equation}
\begin{equation}\label{Xi2}
\begin{split}
\Xi^g(p)&=\frac{1}{N} \int \frac{d^3 k}{(2 \pi)^3}\mathbbm{1}S_F(k)\Gamma(k;p)\Delta_{\varphi}(p-k)+\\
&\!\!\!\!\!\!\!\!\!\!\!\!\!\!\!-\frac{1}{N} \int \frac{d^3 k}{(2 \pi)^3} \mathbbm{1} \textrm{Tr} \left[S_F(k)\Gamma(k;q=0) \Delta_{\varphi}(q=0) \right].
\end{split}
\end{equation}

In order to find the full fermion propagator, it is convenient to decompose this Green function into its irreducible parts. Hence, we adopt the following ansatz
\begin{equation}
S_F^{-1}(p)= -\gamma^{\mu}p_{\mu}A(p)+\Sigma(p),
\label{fullfermion}
\end{equation}
where $A(p)$ yields the wavefunction renormalization and $\Sigma(p)$ is called mass function. Note that $\Sigma(p)\neq 0$ implies a dynamical mass generation \cite{appelquist}. In order to obtain the mass function, we take the trace operation in both sides of Eq. \eqref{sdefermion}.  Thereafter, we substitute Eqs. \eqref{fermionfree}, \eqref{Xi1}, \eqref{Xi2} and \eqref{fullfermion}, within the rainbow approximation, $\Gamma^{\nu}(p;k) \rightarrow \gamma^{\nu}$ e $\Gamma(p;k) \rightarrow \mathbbm{1}$, we obtain
\begin{equation}
\begin{split}
\Sigma(p)&= \frac{\lambda}{N}\int\frac{d^3 k}{(2 \pi)^3}  \frac{\delta^{\mu \nu} \Sigma(k)}{k^2 A^2(k)+\Sigma^2(k)}\Delta_{\mu \nu}(p-k)\\ 
&+ \frac{1}{N} \int \frac{d^3 k}{(2 \pi)^3} \frac{\Sigma(k)}{k^2 A^2(k)+ \Sigma^2(k)}\Delta_{\varphi}(p-k)\\
& - \Delta_{\varphi}(0)\int \frac{d^3 k}{(2 \pi)^3} \frac{\Sigma(k)}{k^2 A^2(k)+ \Sigma^2(k)}.
\label{massfunction}
\end{split}
\end{equation}  
On the other hand, for calculating the wavefuntion renormalization, we multiply Eq. \eqref{sdefermion} by $\slashed{p}$ and, after calculating the trace operation, we obtain 
\begin{equation}
\begin{split}
A(p)&=\!1\! -\!\frac{\lambda}{N p^2}\!\!\int\!\! \frac{d^3 k}{(2 \pi)^3} \left(\delta^{\beta \mu}\delta^{\alpha\nu}-\delta^{\beta \alpha}\delta^{\mu \nu} \right. \\ & \left. +\delta^{\beta \nu}\delta^{\mu \alpha}\right) \frac{p_{\beta}k_{\alpha}A(k)}{k^2A^2(k)+\Sigma^2(k)}\Delta_{\mu \nu}(p-k) \\& +\! \frac{1}{N p^2}\!\!\int\!\! \frac{d^3 k}{(2\pi)^3}  \frac{\delta^{\beta \alpha}p_{\beta}k_{\alpha}A(k)}{k^2 A^2(k)+\Sigma^2(k)}\Delta_{\varphi}(p-k).
\label{wavefunction}
\end{split}
\end{equation}

It is clear that Eqs. \eqref{massfunction} and \eqref{wavefunction} are a coupled set of equations for $\Sigma(p)$ and $A(p)$. However, in the dominant order $1/N$, we may replace $A(p)\rightarrow 1$ in Eq. \eqref{massfunction}. This is also in agreement with the rainbow approximation. 

Next, we use the Matsubara formalism for introducing temperature effects into Eq. \eqref{massfunction}. In the imaginary-time formalism, the main step is to change the time-component integrals into a sum over Matsubara frequencies \cite{matsubara}, i.e, we must perform
\begin{equation}
\int \frac{d k_0}{2\pi} f(k_0,\textbf{k},\beta) \rightarrow \frac{1}{\beta}  \sum_{n=-\infty}^{\infty} f(n, \textbf{k},\beta), 
\end{equation}
with
\begin{equation}
k_{\mu}=(k_0,\textbf{k})  \rightarrow (\omega_n,\textbf{k}), 
\end{equation}
and
\begin{eqnarray*}
\omega_n = \begin{cases}
\frac{(2n+1)\pi}{\beta} \,\, \textrm{for fermions},\\
\frac{2n\pi}{\beta} \hspace{0.7cm} \textrm{for bosons},
\end{cases}
 \end{eqnarray*}
where $\omega_{n}$ are the Matsubara frequencies, $T=\beta^{-1}$ is the temperature of the thermal bath, and $(m, n)$ are the vibration modes related to  each momentum component. Note that we are considering the Boltzmann constant as $k_{B}= 1$, in the natural system of units. After including the thermal bath in Eq.~(\ref{massfunction}), we have
\begin{equation}
\begin{split}
\!\!\!\Sigma_m(\textbf{p},T)&= C_1 \sum \!\!\!\!\!\!\!\! \int \frac{\Sigma_n(\textbf{\textbf{k}, T})}{(2n+1)^2 \pi^2 T^2 + \textbf{k}^2 +\Sigma^2_{n}(\textbf{k},T)} \\ & \!\!\!\! \times \frac{1}{\sqrt{4 (n-m)^2 \pi^2 T^2 +(\textbf{p}-\textbf{k})^2}} \,\,+ \\
&+C_2 \sum \!\!\!\!\!\!\!\! \int \frac{\Sigma_n(\textbf{k},T)}{(2n+1)^2 \pi^2 T^2 + \textbf{k}^2 +\Sigma^2_{n}(\textbf{k},T)}  \\ 
& \!\!\!\! \times \frac{1}{(g g_0)^{-1}+\sqrt{4 (n-m)^2 \pi^2 T^2 +(\textbf{p}-\textbf{k})^2}},
\label{masstemperature}
\end{split}
\end{equation}
where
\begin{equation}
\sum \!\!\!\!\!\!\!\! \int  \rightarrow T\sum_{n = -\infty}^{\infty} \int \frac{d^2 k}{(2 \pi)^2}, 
\end{equation}
\begin{equation}
C_1 = \frac{2 \lambda }{\left(2+\frac{\lambda}{8}\right)N},
\end{equation}
\begin{equation}
C_2  = \frac{1}{g_0 N}.
\end{equation}

From now on, we assume that $\Sigma_m(\textbf{p},T)=\Sigma(\textbf{p},T)$ and, therefore, the mass function only depends on the dominant vibrational mode $m=0$. The standard procedure for calculating the mass function is to convert the integral equation into a differential equation for $\Sigma(\textbf{p},T)$. It turns out that there exist two main operations, namely, the sum over $n$ and the angular integral that must be performed before finding this result. We shall explore some approximations for these operations.

Before performing further approximations, let us show that the sum over the vibrational modes $n$ is convergent. The convergence of this sum for PQED at finite temperature has been shown in \cite{PQEDT}. In Eq.~(\ref{masstemperature}), the sum over $n$ may be written as
\begin{equation}
U=\sum_{n=-\infty}^{\infty}\frac{1}{(2n+1)^2+A^2}\frac{1}{B+\sqrt{n^2+C^2}},
\label{GNconv}
\end{equation}
where we conclude that $(A^2,C^2)>0$ and, because we consider $g>0$ \cite{gross,van}, we also find that $B>0$. 

Next, we apply the Cauchy integral test \cite{arfken}, where the sum is written as
\begin{equation}
U=2 \sum_{n=1}^{\infty}u^{(+)}_n + u_0,
\label{GN2}
\end{equation}
with $u^{(+)}_n = (u_n+u_{-n})/2$. The test imposes that whether $u(x)=u_x$ is positive, continuous, and decreasing in the interval $[1, \infty]$, hence, the integral over $u(x)$ is finite and, therefore, convergence is derived. This is exactly our case and one may easily conclude that
\begin{equation}
u(x)>0, \,\,\,\,\, \forall x \mid  x\,\in  [1,\infty],
\end{equation}
\begin{equation}
u'(x)<0,
\end{equation}
\begin{equation}
\lim_{x\rightarrow a} u(x)  = u(a), \,\, \forall a \mid \,\, a \in [1,\infty].
\end{equation} 
Therefore, Eq.~\eqref{GN2} is a convergent series for all $n$.

\section{\textbf{STATIC REGIME} $p_0 =0$}
The static regime, also called instantaneous-exchange approximation \cite{exchange}, consists of neglecting all of the time-components of the bosonic-field propagator in Eq. \eqref{massfunction}, with the consideration that the interaction vertex being only $\gamma^0$ in Eq. \eqref{Xi1}. This is usually realistic in cases where the electron velocity is much less than the light velocity. In these systems, it is possible to show that charged particles interacts through the Coulomb potential. This is given by the Fourier transform of the gauge-field propagator in Eq.~(\ref{photonbare}) after we perform $p_\mu=(p_0,\textbf{p}) \rightarrow (0,\textbf{p})$, where we will use the following notation $(0,\textbf{p})\equiv (0,P)$. Next, let us consider both zero and finite temperature cases for calculating $\Sigma(P,T)$.

\subsection{ \textbf{Zero Temperature Case}}

After considering the static regime with $T\rightarrow 0$ in Eq. \eqref{masstemperature}, we find
\begin{equation}
\begin{split}
\Sigma(P)&=\frac{C_1}{2} \int \frac{d^3k}{(2\pi)^3} \frac{\Sigma(K)}{k_0^2+K^2+\Sigma^2(K)}\Delta(P-K) \\ 
&+C_2\!\! \int \frac{d^3k}{(2\pi)^3} \frac{\Sigma(K)}{k_0^2+K^2+\Sigma^2(K)}\Delta_\varphi(P-K), \label{SP0}
\end{split}
\end{equation}
where $d^3k=dk_0 d^2K$,
\begin{equation}
\Delta_\varphi(P-K)=\frac{1}{(gg_0)^{-1}+\sqrt{(P-K)^2}},
\end{equation} 
and
\begin{equation}
\Delta(P-K)=\frac{1}{\sqrt{(P-K)^2}}.
\end{equation}

The integral over $k_0$ in Eq. \eqref{SP0} is easily solved and the angular integral may be solved by using the identity
\begin{equation}
\int_0^{2\pi} d\theta \Delta_\varphi(P-K)=\frac{4K(x_0)}{\sqrt{(P-K)^2}}+O[(gg_0)^{-2}], \label{angT}
\end{equation}
where $K(x_0)$ is the complete elliptic integral of the first kind with $x_0\equiv -4 PK/(P-K)^2$. Note that the angular integral for the gauge-field term is very similar and that we will neglect the terms in order of $O[(gg_0)^{-2}]$. Replacing Eq.~(\ref{angT}) in Eq.~(\ref{SP0}), we find
\begin{equation}
\Sigma(P)=4\left(\frac{C_1}{2}+C_2\right) \int_0^\Lambda  \frac{dK}{(2\pi)^2}  \frac{K \,\Sigma(K) {\cal K}(K,P)}{2\sqrt{K^2+\Sigma^2(K)}}, \label{SigP}
\end{equation}
where the kernel ${\cal K}(K,P)$ reads
\begin{equation}
{\cal K}(K,P)=\frac{K(x_0)}{\sqrt{(P-K)^2}}.
\end{equation}

Next, we approximate the kernel for its lowest order terms, namely,
\begin{equation}
{\cal K}(K,P) \approx \frac{\pi}{2P} \Theta(P-K)+\frac{\pi}{2K} \Theta(K-P). \label{kernelApp}
\end{equation}
Note that $\Theta(P-K)$ is the standard step function. Using Eq.~(\ref{kernelApp}) in Eq.~(\ref{SigP}), we have
\begin{equation}
\begin{split}
\Sigma(P)&=\frac{(C_1/2+C_2)}{4\pi} \left[\int_0^P K dK \frac{\Sigma(K)}{P\sqrt{K^2+\Sigma^2(K)}} \right. \\
& \left. +\int_P^\Lambda  dK \frac{\Sigma(K)}{\sqrt{K^2+\Sigma^2(K)}}\right].
 \label{SigP2}
\end{split}
\end{equation}

On the other hand, by taking derivatives in respect to $P$ in both sides of Eq.~(\ref{SigP2}), we obtain  a differential equation for the mass function, given by
\begin{equation}
P^2 \Sigma''(P)+2P \Sigma'(P)+\frac{N_c}{4N} \frac{P \Sigma(P)}{\sqrt{P^2+\Sigma^2(P)}}=0, \label{EulerT0}
\end{equation}
where the critical number of fermions is given by
\begin{equation}
N_c=\frac{1}{\pi}\left(\frac{C_1}{2}+C_2\right)=\frac{ \lambda }{\pi\left(2+\frac{\lambda}{8}\right)}+\frac{4}{\pi}. \label{Nc0}
\end{equation}
Eq. \eqref{EulerT0} is also supplemented by two equations, namely,
\begin{equation}
\lim_{P\rightarrow 0} P^2 \Sigma'(P)=0,
\end{equation}
which provides the infrared behavior and
\begin{equation}
\lim_{P\rightarrow \Lambda} \left[ P \Sigma'(P)+\Sigma(P) \right]=0,
\end{equation}
which shows that the mass function is expected to vanish at large $P$. 

Using the approximation $P/\sqrt{P^2+\Sigma^2(P)} \approx 1$ in Eq. \eqref{EulerT0}, which holds for large-external momentum, we find the analytical solution 
\begin{equation}
\Sigma(P)=A_1 P^\gamma+A_2 P^{\gamma^*}, \label{SolEuler}
\end{equation}
where $\gamma=-1/2+i/2 \sqrt{N_c/N-1}$ and $(A_1,A_2)$ are arbitrary constants. In the dynamical regime, it has been show that the dynamical mass generation only occurs for $N<N_c$ \cite{GNT=0}. The same conclusion  form Eq. \eqref{Nc0} is also obtained from the static regime. Using Eq. \eqref{Nc0} in $\gamma$,  $\lambda = 4 \pi N \alpha_c$, and solving $N_c(\alpha_c)/N=1$ for $\alpha_c$, we find the critical fine-structure constant, given by
\begin{equation}
\alpha_c =  \frac{ \frac{16}{N \pi} - 4}{N \pi - 12 },
\end{equation}
in terms of fine-structure constant the mass generation condition is $\alpha > \alpha_c$. 

\subsection{ \textbf{Finite Temperature}}

In order to describe temperature effects, we include the Matsubara frequencies in Eq. \eqref{SP0}. Thereafter, we solve the angular integral using Eq.~(\ref{angT}) while for the Matsubara sum we use the identity
\begin{equation}
T\sum_n \frac{1}{\omega^2_n+\epsilon^2_k}=\frac{1}{2\epsilon_k}\left[1-2n_F(\epsilon_k)\right],
\end{equation}
where $\epsilon_K \equiv \sqrt{K^2+\Sigma^2(K)}$ and
\begin{equation}
n_F(\epsilon_K)=\frac{1}{e^{\beta \epsilon_K}+1}
\end{equation}
is the Fermi-Dirac distribution. After using these properties, we find
\begin{equation}
\!\!\!\!\!\Sigma(P,T)\!=\!4 \left(\!\frac{C_1}{2}\!+\!C_2\!\right) \int_0^\Lambda \!\!\!\!dK \frac{K}{(2\pi)^2} \frac{ \Sigma(K){\cal K}_\beta (K,P)}{2\sqrt{K^2+\Sigma^2(K)}}, \label{IEST}
\end{equation}
where the temperature-dependent kernel ${\cal K}_\beta(K,P)$ reads
\begin{equation}
{\cal K}_\beta(K,P)=\frac{K(x_0)}{\sqrt{(P-K)^2}}\left[1-2n_F(\epsilon_K)\right].
\end{equation}

Next, following the same steps as in the previous section, we find the differential equation for the mass function, namely,
\begin{equation}
\!\!P^2 \Sigma''(P)\!+\!2P \Sigma'(P)\!+\!\frac{N_c h_\beta(P)}{4N}\frac{P \Sigma(P)}{\sqrt{P^2+\Sigma^2(P)}}=0, \label{EulerT}
\end{equation}
where
\begin{equation}
h_\beta(P)=\left[1-2n_F(\epsilon_P)\right].
\end{equation}
The linearized version of Eq.~(\ref{EulerT}) may be obtained at large-external momentum and by doing $P\rightarrow \Lambda$ in $h_\beta(P)$. This allow us to obtain a temperature-dependent critical number $N_c(T)$, given by
\begin{equation}
N_c(T)=N_c(0) \left[1-\frac{2}{e^{\Lambda/T}+1}\right], \label{NcT}
\end{equation}
where $N_c(0)\equiv N_c=(C_1/2+C_2)/\pi$ is given by Eq. \eqref{Nc0} in terms of $\lambda$. On the other hand, for large-$T$, we obtain $N_c(T)\rightarrow 0$, which implies that there exist no dynamical mass generation for any $\lambda$. 

\begin{figure}[H]
\centering
\includegraphics[scale=0.9]{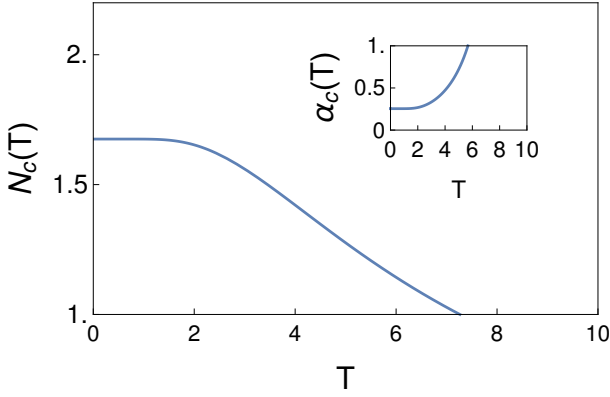}
\caption{The critical parameters for the dynamical mass generation. We plot both Eq.~(\ref{NcT}) (the common line) and Eq.~(\ref{alphacT}) (the inset) with $\Lambda=10$ (units of energy). For the common line we also use $\lambda=3.0$ while for the inset we use $N=2$. Note that for small enough temperatures, the critical coupling constants remain almost unchanged.} \label{Fig4}
\end{figure}

\begin{figure}[H]
\includegraphics[scale=0.9]
{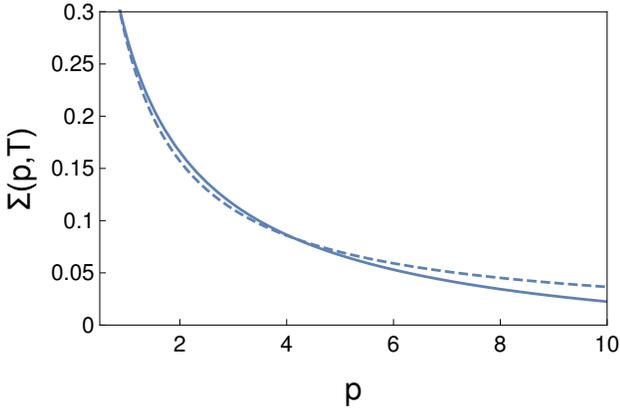}
\caption{The analytical and numerical solutions for $\Sigma(P,T)$. The dashed line is the numerical solution of the integral equation given by Eq. \eqref{IEST} with $N=1$. The common line is the analytical solution given by Eq. \eqref{SolEulerT} with $N=1$ and $A_1(T)=A_2(T)^*=0.13+0.05i$. We have used $\Lambda=10$ (units of energy), $\lambda=3.0$ and $T=0.1$ (units of $\Lambda$) for all of the curves. The analytical solution  is in good agreement with the numerical results.} \label{Fig5}
\centering
\end{figure}

In the large-external momentum, the solution of Eq. \eqref{EulerT} is given by Eq. \eqref{SolEuler}, after we replace $N_c \rightarrow N_c(T)$. Hence, we obtain a temperature-dependent exponent $\gamma(T)=-1/2+ i/2\sqrt{N_c(T)/N-1}$. Therefore, we find
\begin{equation}
\Sigma(P,T)=A_1(T) P^{\gamma(T)}+A_2(T) P^{\gamma^*(T)}, \label{SolEulerT}
\end{equation}
where the arbitrary constants $(A_1,A_2)$ may be dependent on the temperature. At the critical point $N=N_c(T)$, we have $\gamma(T)=-1/2$. This critical point also may be described in terms of $\alpha$, using Eq.~(\ref{Nc0}) and that $e^2=4\pi\alpha= \lambda N$. Indeed, after a simple calculation, we obtain a temperature-dependent critical coupling constant $\alpha_c(T)$, namely,
\begin{equation}
\alpha_c(T)\!=-\frac{4 \left[4\left(1-e^{\Lambda/T}\right) + N\pi\left(1+e^{\frac{\Lambda}{T}}\right)\right]}{ N \pi \left[12\left(1-e^{\Lambda/T}\right) +N \pi\left(1+e^{\frac{\Lambda}{T}}\right)\right]}. \label{alphacT}
\end{equation}
Because the ratio $N_c(T)/N$ is a monotonically increasing function of $\alpha$, it follows that the mass generation only occurs for $\alpha>\alpha_c(T)$. In Fig.~\ref{Fig4}, we plot both $N_c(T)$ and $\alpha_c(T)$ and show that whether the temperature increases, hence, the system quickly goes to a gapless phase.
We also may find numerical results (see App.~A for more details) of the integral equation for $\Sigma(P,T)$ in Eq.~(\ref{IEST}). In Fig.~\ref{Fig5}, we plot these results and compare this with our analytical solution given by Eq.~(\ref{SolEulerT}).

\section{\textbf{DYNAMICAL REGIME}  $p_0 \neq 0$}

In this section we recover the retardation effects by assuming $p_0 \neq 0$ in Eq. \eqref{massfunction}. For $T=0$, this regime has been discussed in several contexts.

After a simple changing of variables, the angular integral of the PQED sector (see details in App. B) reads
\begin{equation}
\mathcal{I}_n^{PQED}(p,k;T)=\frac{4}{w_{p,k}^n}K\left[x_n\right],
\end{equation}
where,
\begin{equation}
w_{p,k}^n = \sqrt{4n^2\pi^2T^2 +(p-k)^2} \label{omega},
\end{equation}
and
\begin{equation}
x_n = -\frac{4pk}{4n^2\pi^2T^2 +(p-k)^2} \label{x}. 
\end{equation}
Similarly, for the GN sector, we find
\begin{equation}
\mathcal{I}_n^{GN}(p,k,T)\! =\!\frac{4}{w_{p,k}^n}\left[K(x_n)-\frac{\Pi(y_n,x_n)}{1-g^2 g_0^{2}\left(w_{p,k}^n\right)^2}\right]
\end{equation} 
with
\begin{equation}
y_n=-\frac{4pk}{4n^2\pi^2T^2 + (p-k)^2 -\frac{1}{g^2 g_0^{2}}},
\label{y}
\end{equation}
where $K(x_n)$ and $\Pi(y_n,x_n)$ are the elliptic integral of the first and third kind, respectively. In particular, note that for  $g=0$, we have that $\Pi(0,x_n) = K(x_n)$, hence, the kernel of the GN sector vanishes, as expected. The integral representation for these functions are
\begin{align}
K(x_n)= \int_0^{\pi/2} \!\!\!\! d\theta\frac{1}{\sqrt{1-x_n \sin^2(\theta)}}
\label{ellipticK}
\end{align}
 and
\begin{equation}
\Pi(y_n,x_n)=\!\! \int_0^{\pi/2} \!\!\!\! d\theta\frac{1}{[1-y_n \sin^2(\theta)]\sqrt{1-x_n \sin^2(\theta)}}.
\label{ellipticPi}
\end{equation}

The resulting kernel, after using the angular integral, is a very complicated function of $n$. This preclude us of solving the whole Matsubara sum, as we have did in the static regime. Therefore, we shall investigate the lowest order contribution, provided by the zero-mode $n=0$ approximation.  

\section{\textbf{Zero-mode approximation}}

In this section we consider the fundamental vibrational mode $n=0$ in Eq. \eqref{masstemperature}. Using Eqs.~\eqref{ellipticK} and \eqref{ellipticPi}, we find   
\begin{equation}
\begin{split}
\Sigma(p,T)&=\frac{C_1}{\pi^2}\!\! \int_0^{\infty} \!\!\!\!\!\!dk \frac
{k K(x_0)}{\|p-k \|} \frac{T \Sigma(k,T)}{\pi^2 T^2 + k^2 +\Sigma^2(k,T)}\\
& + \frac{C_2}{\pi^2}\!\!\int_0^{\infty}\!\!\!\!\!\!dk \frac{k}{\|p-k \|} \frac{T\Sigma(k,T)}{\pi^2 T^2 + k^2 + \Sigma^2(k,T)} \\
& \times \left[K(x_0)-\frac{\Pi(y_0,x_0)}{1-g^2 g_0^{2}(p-k)^2}\right] 
\label{zeromode}
\end{split}
\end{equation}

For calculating an analytical solution of Eq.~(\ref{zeromode}), we write the elliptic integral as the hypergeometrics Appell $F_1$. Because $F_1$ has an expansion in its parameters, we may find a simplified version of the kernel (see App. C for more details). Hence,  Eq. \eqref{zeromode} yields 
\begin{equation}
\begin{split}
\Sigma(p)&=\frac{C_1 T}{2 \pi}\int_0^{\Lambda} \!\!dk f(k) \left[\frac{\Theta(p-k)}{p}+\frac{\Theta(k-p)}{k}\right] \\ & + \frac{C_2 T}{2 \pi}\int_0^{\Lambda} \!\!\! dk f(k) \left[\frac{\Theta(p-k)}{p}+\frac{\Theta(k-p) k}{k^2 -\frac{1}{(g g_0)^2}}\right], \label{auxST}
\end{split}
\end{equation}
where $f(k)\equiv k \Sigma(k)/\left[(\pi T)^2+k^2+\Sigma^2(k)\right]$. By taking derivatives in respect to $p$ in both sides of Eq.~(\ref{auxST}), we find 
\begin{equation}
\begin{split}
\Sigma'(p)+\frac{b \Sigma(p)}{(\pi T)^2+p^2+\Sigma^2(p)}= -\frac{a}{p^2}\int_0^{p}dk f(k), \label{Sigmaprime}
\end{split}
\end{equation} 
with $a=(C_1+C_2)T/2 \pi$ and $b=C_2 T/2 \pi$, both constants with dimension of mass. Next, we neglect the nonlinear terms by using $(\pi T)^2+p^2+\Sigma^2(p) \approx (\pi T)^2+p^2$, which is expected to be reasonable at large $p$ \cite{cutinfrared, PQEDT}. After deriving Eq.~(\ref{Sigmaprime}) in respect to $p$, we obtain the differential equation
\begin{equation}
\begin{split}
p^3 & \Sigma''(p) +\left[2p^2+\frac{p^3 b}{(\pi T)^2+p^2}\right]\Sigma'(p) \\
& +\left[\frac{2 p^2 b}{(\pi T)^2+p^2}-\frac{2 p^4 b}{\left((\pi T)^2+p^2\right)^2}+a\right]\Sigma(p)=0,
\label{diff}
\end{split}
\end{equation} 
which is supplemented by two conditions for the infrared and ultraviolet regimes. These are given by
\begin{equation}
\lim_{p \rightarrow \Lambda} \Sigma(p)=0
\end{equation}
and
\begin{equation}
\lim_{p \rightarrow 0} p^2 \left[\frac{d \Sigma(p)}{d p} +\frac{b \Sigma(p)}{(\pi T)^2+p^2}\right] = 0. \label{IR}
\end{equation}

Unfortunately, Eq.~(\ref{diff}) has not an analytical solution for arbitrary values of the set $(p,T)$ of variables. Nevertheless, we may consider a possible linearized version for  large external momentum.

\subsection{\textbf{Large-external-momentum  approximation}}

\begin{figure}[H]
\includegraphics[scale=0.6]{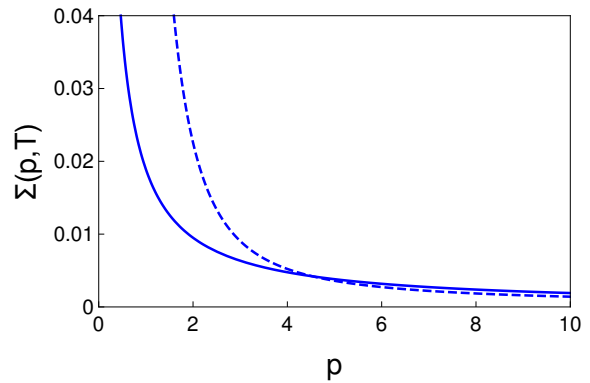}
\centering
\caption{The analytical and numerical solutions for $\Sigma(p,T)$ in the zero-mode approximation. The dashed line is the numerical solution of the integral equation given by Eq.~(\ref{zeromode}) with $N=2$. The common line is the analytical solution given by Eq.~(\ref{anazero}) with $N=2$ and $A_1(T)=0.6$ (the best fit for large-$p$). We have used $\Lambda=10$ (units of energy), $\lambda=3.0$ and $T=0.1$ (units of $\Lambda$) for all of the curves. Note that the analytical solution  is in good agreement with the numerical results only at large-$p$ limit, as expected.} \label{Fig6}
\end{figure}

In this section we investigate the linearized version of Eq.~(\ref{diff}) by assuming $p\gg T,\Sigma(p)$. In this case, the behavior of the generated mass is given by
\begin{equation}
p^3  \Sigma''(p) +\left[2p^2+pb\right]\Sigma'(p)  +a\Sigma(p)=0,
\label{diffp}
\end{equation} 
which yields 
\begin{align}
\!\!\Sigma(p)\!=\!A_1 \frac{b \,_1F_1\left[1-\frac{a}{b},2,\frac{b}{p}\right]}{p}\! + A_2  G^{2\, 0}_{1\,2} \left[ -\frac{b}{p} \big{|}
\begin{array}{c l}
\frac{a+b}{b}\\
0,1
\end{array} \right], \label{anazero}
\end{align}
where $_1F_1$ is the confluent hypergeometric function and G is the Meijer G-function \cite{handbook}. The sign of this Meijer G-function changes for different values of the external momentum. Because we have not observed such behavior in the numerical results of Eq.~(\ref{zeromode}), from now on, we shall take $A_2=0$ for the sake of agreement with the integral equation. In Fig.~(\ref{Fig6}), we plot our analytical solution in Eq.~(\ref{anazero}) and compare this with the numerical results of the integral equation in Eq.~(\ref{zeromode}).

 \vspace{-1.1cm}
\subsection{\textbf{Zero-external-momentum approximation}}

In this section, we consider the limit $\Sigma(p=0,T)=m(T)$ in Eq. \eqref{masstemperature}, using the zero-frequency mode. In this case, the angular integral provides a constant factor of $2 \pi$. Hence, we find
\begin{equation}
\begin{split}
1&= \frac{1}{4 \pi}\int_{0}^{\Lambda}\!\!\!\! dk \,\, \frac{1}{\pi^2 T^2 + k^2 +m^2(T)} \\ & \times\left\lbrace C_1 T + \frac{C_2 T k}{(g g_0)^{-1}+k}\right\rbrace. 
\end{split}
\end{equation}
After calculating the integral over $k$, we have

\begin{widetext}
\begin{equation}
\begin{split}
1&= \frac{C_1 T}{2 \pi} \frac{\arctan \left(\frac{\Lambda}{\sqrt{\pi^2 T^2 + m^2(T)}}\right)}{ \sqrt{m^2(T)+  \pi^2T^2}} + \frac{C_2 T}{2 \pi} \left\lbrace \frac{\sqrt{\pi^2 T^2 + m^2(T)}\arctan \left(\frac{\Lambda}{\sqrt{\pi^2 T^2 + m^2(T)}}\right)}{\left[\pi^2 T^2 + m^2(T) + (\frac{1}{g g_0})^2\right]} \right. \\ & \left. + \frac{g g_0 \ln\left[\frac{\pi^2 T^2 + m^2(T)+\Lambda^2}{(\pi^2 T^2 + m^2(T))(1 +g g_0 \Lambda)^2}\right]}{2\left[(g g_0)^2(\pi^2 T^2 +m^2(T))+1\right]}\right\rbrace.
\label{m(T)}
\end{split}
\end{equation}
\end{widetext}

Equation \eqref{m(T)} is our gap equation for the dynamical mass as a function of temperature and the coupling constants . Similarly to the BCS model, we define the critical temperature as the point $T=T_c$ in which $m$ vanishes. Here, nevertheless, we define that $C_1 T$ and $C_ 2 T$ are fixed in order to be in agreement with the fact that $m\neq 0$ for $T=0$. 
\begin{figure}[H]
\centering
\includegraphics[width=8cm, height=4cm]{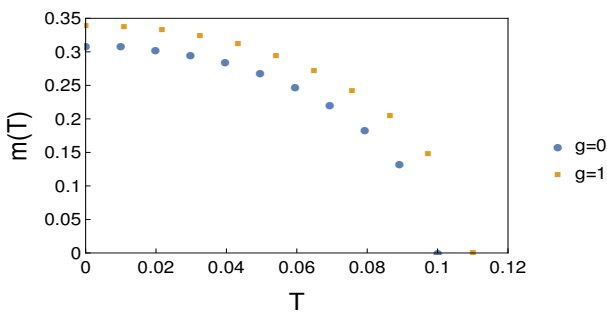}
\caption{The numerical result of the mass function given by Eq. \eqref{m(T)}. We use $\Lambda =10$, $N=2$ and $\lambda=3$ in the blue points with $g = 0$, and in the square point $g=1$. Where, we see the aument of the critical temperature due at the GN-coupling.}
\label{mT}
\end{figure}

\begin{figure}[H]
\centering
\includegraphics[width=8cm, height=5cm]{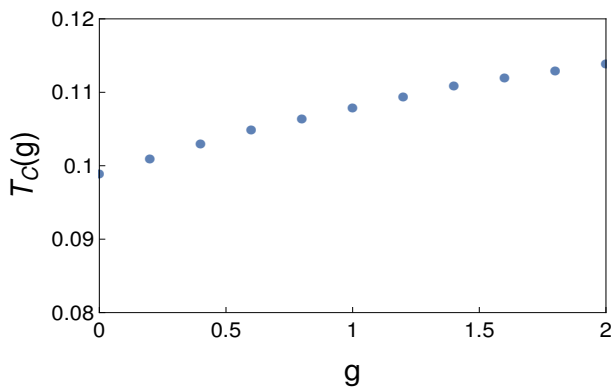}
\caption{Behaviour of critical temperature for different values of the GN coupling constant given by Eq. \eqref{m(T)} numerically treated. Where we use $\Lambda=10$, $N=2$ and $\lambda=3$.}
\label{T(g)}
\end{figure}

In Fig. \ref{mT}, we plot the numerical solutions of $m(T)$ for both $g=0$ and $g=1.0$, which proves that the GN interaction increases the gapped phase.  Furthermore, in Fig. \ref{T(g)}, we show that the GN interaction increases the critical temperature, as expected.

\section{\textbf{Summary and Outlook}}
In this work, we investigate the influence of temperature on the dynamical mass generation for fermions in PQED with a Gross-Neveu interaction. In order to do so, we used the Schwinger-Dyson equation for the electron in the dominant order $1/N$, negleting quantum corrections to the vertices at the finite temperature. In the static regime, we were able to solve the whole Matsubara sum. Thereafter, we calculated the critical coupling constant $\alpha_c$ and the critical numbers of fermions $N_c$ as a function of $T$ and the cutoff $\Lambda$. From these results, we concluded that the system goes to a gapless phase whether we increase the ratio $T/\Lambda$. Furthermore, we obtained an analytical solution for the mass function $\Sigma(p)$, which is in good agreement with our numerical results.

In the dynamical regime, we used the zero-mode approximation in the sum over the Matsubara frequencies, which allowed us to calculate the analytical solution $\Sigma(p)$. This function agrees with our numerical results for large-exteral momentum, while for small-external momentum some deviations have been found. Moreover, we found a gap equation that provides the value of the mass function in the zero-external momentum approximation, i.e, $\Sigma(p=0,T) \rightarrow m(T)$. In this case, the numerical tests show that the critical temperature increases as we increase the strength of the fermionic self-interaction. 

Several generalizations of this work may be perfomed. For instance, one would investigate the main effects of vertices quantum corrections into the critical parameters for the mass generation. In particular, it would be relevant to find a way to increase the gapped phase for finite temperature. In this phase, the competition between   $\alpha$ and $g$ yields a tunneable mass which may be relevant for applications in two-dimensional materials \cite{PQEDGap}. 

\section*{\textbf{ACKNOWLEDGMENTS}}
L. F.  is partially supported by Coordena\c{c}\~{a}o de Aperfei\c{c}oamento de Pessoal de N\'{i}vel Superior Brasil (CAPES), finance code 001. V. S. A. and L. O. N. are partially supported by Conselho Nacional de Desenvolvimento Cient\'{i}fico e Tecnol\'{o}gico (CNPq) and by CAPES/NUFFIC, finance code 0112. F. P. acknowledge the financial support from DIUFRO Grant DI20-0005 of the Direcci\'{o}n de Investigaci\'{o}n y Desarrollo, Universidad de La Frontera. 

\appendix
\numberwithin{equation}{section}
\textbf{\section{\textbf{The Numerical Results}}}

In this Appendix we briefly review the main steps in order to find our numerical results. For the sake of simplicity, note that Eq. \eqref{masstemperature} is a kind of Fredholm integral equation of first kind, defined by
\begin{equation}
Z(x)=\int_a^{b}\! dy\,\, \mathcal{K}\textbf{(}x,y,Z(y)\textbf{)}Z(y), \,\,\,\,\, (a\leq x \leq b)
\label{fredholm}
\end{equation}
where $\mathcal{K}\textbf{(}x,y,Z(y)\textbf{)}$ is the kernel of the integral equation and $Z(x)$ is an unknown function. Throughout this work we convert this into a differential equation, allowing us to find analytical solutions for each approximation, namely, Eq.~\eqref{SolEulerT} in the static approximation and Eq.~\eqref{anazero} in the zero-mode approximation. These solutions, nevertheless, are not expected to well describe the behavior of the mass function for small external momentum.
 
For properly describing the behavior of $\Sigma(p)$ for any $p$, we may find numerical results of both  Eq. \eqref{IEST} and Eq. \eqref{auxST}. It is easy to conclude that these equations are essentially a kind of Fredholm integral equation whether we adjust the parameters, variables, and kernel.  Next, let us summarize our main steps. First, we use the trapezoidal quadrature rule for calculating the integral over the kernel, hence,
\begin{eqnarray}
\int_a^{b}dy \mathcal{K}(x,y,Z)Z(y)  & \rightarrow &  \frac{h}{2} \sum_{i=1}^{M-1} [\mathcal{K}(y_i,x,Z_i) Z_i  \nonumber \\  &+&\mathcal{K}(y_{i+1},x,Z_{i+1})Z_{i+1}], \label{trap}
\end{eqnarray} 
where $h$ is the size of each interval and $y_i$ is the iterative variable (plays the role of loop momentum $k$). Setting $M = 300$ as the numbers intervals, hence, $h=(a-b)/(M-1)$, with $a=10$ (plays the role of the $\Lambda$ cutoff), and $b=10^{-3}$ in order to avoid divergences. Furthermore, note that $10^{-3}$ $\leq$ $(x,y_i)$ $\leq$ $10$.  Using Eq.~(\ref{trap}) in Eq.~(\ref{fredholm}), we obtain a set of algebraic equations for all of $Z_i$ from $x=b$ to $x=a$. Finally, after solving this set of equations, we find Fig.~\eqref{Fig5} and Fig.~\eqref{Fig6}.

\textbf{\section{\textbf{Angular Integral }}}

The mass function given by Eq. \eqref{masstemperature} has two main contributions, generated by PQED and GN interactions. In the polar system of coordinates, it yields  
\begin{widetext}
\begin{equation}
\begin{split}
&\Sigma(p)=C_1 T \sum_{n=-\infty}^{\infty}\! \int_0^{\infty}\!\! \frac{d k}{(2 \pi)^2} \frac{k\,\, \Sigma(k)}{(2n +1)^2 \pi^2 T^2+k^2+\Sigma^2(k)}\int_0^{2 \pi}\!\!d \theta\frac{1}{\sqrt{4n^2 \pi^2 T^2 + p^2+k^2-2 k p \cos(\theta)}} \\
& + C_2 T \sum_{n=-\infty}^{\infty}\! \int_0^{\infty}\!\! \frac{d k}{(2 \pi)^2} \frac{k \,\,\Sigma(k)}{(2n +1)^2 \pi^2 T^2+k^2+\Sigma^2(k)} \int_0^{2 \pi}\!\!d \theta\frac{1}{\frac{1}{g g_0}+\sqrt{4n^2 \pi^2 T^2 + p^2+k^2-2 k p \cos(\theta)}}. 
\end{split}
\end{equation}
\end{widetext}
After we use the identity $\cos(\theta) = 1-2 \sin^2(\theta/2)$ and a change of variable ($\theta$ $\rightarrow$ $2 \theta$) in the angle, it is possible to write the PQED angular integral as
\begin{equation}
\begin{split}
\mathcal{I}^{PQED}_n &= \frac{4}{\omega^n_{p,k}}\int_0^{\pi/2} d \theta \frac{1}{\sqrt{1-x_n\,\sin^2(\theta)}}\\
&=\frac{4}{\omega^{n}_{p,k}}K(x_n),
\end{split}
\end{equation}
where $\omega^n_{p,k}$ and $x_n$ are given in Eqs. \eqref{omega} and \eqref{x}, respectively. The angular integral corresponds to an elliptical integral of the first kind, whereas the angular integral of the GN sector reads
\begin{equation}
\mathcal{I}^{GN}_n= \int^{2 \pi}_0 \!\! d \theta\frac{1}{\frac{1}{g g_0}+\omega^n_{p,k}\sqrt{1-x_n \sin^2(\theta)}}. \label{IGN}
\end{equation}
After some algebra, it is possible to write Eq.~(\ref{IGN}) as the sum of three integrals, using the cosine double angle identity ($\cos(\theta) = 1-2 \sin^2(\theta/2)$), therefore, $\mathcal{I}^{GN}_n$ is given by
\begin{align*}
\left(\mathcal{I}^1\right)^{GN}_n&= \frac{2}{g g_0 y_n}\int^{\pi}_0 \! d\theta\frac{1}{1-y_n \sin^2(\theta)}, \\
\left(\mathcal{I}^2\right)^{GN}_n&= \frac{2}{\omega^{n}_{p,k}}\int^{\pi}_0 \! d\theta\frac{1}{\sqrt{1-x_n \sin^2(\theta)}},\\
\left(\mathcal{I}^3\right)^{GN}_n&= -\frac{2}{\omega^{n}_{p,k}}\int^{\pi}_0 \! d\theta\frac{1}{(1-y_n \sin^2(\theta))\sqrt{1-x_n \sin^2(\theta)}},
\end{align*}
where the integral $\left(\mathcal{I}^1\right)^{GN}_n$ vanishes and, after properly writing the integrals $\left(\mathcal{I}^2\right)^{GN}_n$ and $\left(\mathcal{I}^3\right)^{GN}_n$, we obtain 
\begin{equation}
\mathcal{I}^{GN}_n=\frac{4}{\omega^{n}_{p,k}}\left[K(x_n)-\frac{\Pi(y_n,x_n)}{1-g^2 g^{2}_0 (\omega^{n}_{p,k})^2}\right],
\end{equation}
with $y_n$ being defined in Eq.~\eqref{y}. 

Finally, using these results, we find the mass function, namely,
\begin{widetext}
\begin{equation}
\begin{split}
&\Sigma(p)=\frac{C_1T}{(2\pi)^2} \sum_{n=-\infty}^{\infty}\! \int_0^{\infty}\!\! d k \,\,4\,\,\frac{K(x_n)}{\omega^{n}_{p,k}}\frac{k\,\,\Sigma(k)}{(2n +1)^2 \pi^2 T^2+k^2+\Sigma^2(k)} \\
& + \frac{C_2T}{(2\pi)^2} \sum_{n=-\infty}^{\infty}\! \int_0^{\infty}\!\! d k \frac{4}{\omega^{n}_{p,k}}\left[K(x_n)-\frac{\Pi(y_n,x_n)}{1-g^2 g^{2}_0 (\omega^{n}_{p,k})^2}\right] \frac{k\,\,\Sigma(k)}{(2n +1)^2 \pi^2 T^2+k^2+\Sigma^2(k)} . 
\label{masselliptic}
\end{split}
\end{equation}
\end{widetext}

\textbf{\section{\textbf{The Kernel Expansion }}}

The complete elliptic integral of the first and third kinds may be written in terms of hypergeometric function of two variables, for $|x_n|<1$ and $|y_n|<1$ \cite{handbook}, we have
\begin{equation}
K(x_n)=\frac{\pi}{2} F_1\left(\frac{1}{2};\frac{1}{2},1;1,x_n,0\right),
\end{equation}
\begin{equation}
\Pi(y_n,x_n)=\frac{\pi}{2} F_1\left(\frac{1}{2};\frac{1}{2},1;1,x_n,y_n\right).
\end{equation}
The series of the Hypergeometric Appell F1 is given by 
\begin{equation}
F_1(\alpha;\beta,\beta',\gamma;x_n,y_n)=\sum_{l,q=0}^{\infty}\frac{(\alpha)_{l+q}(\beta)_l (\beta')_q}{(\gamma)_{l+q} l! q!}x_n^l y_n^q.
\end{equation} 
The standard procedure in the treatment of the SDEs is to consider two regions (infrared and ultraviolet) in the external momentum. After considering the zero-mode frequency $n=0$, we find
\begin{equation}
2 \pi \frac{F_1\left(\frac{1}{2};\frac{1}{2},1;1,x_0,0\right)}{|p-k|}=\Theta (p - k) \frac{2\pi}{p} + \Theta (k - p) \frac{2 \pi}{k}
\end{equation}
 and
\begin{equation}
\begin{split}
\frac{2 \pi}{|p-k|}\!\!&\left[F_1\left(\frac{1}{2};\frac{1}{2},1;1,x_0,0\right)-\frac{F_1\left(\frac{1}{2};\frac{1}{2},1;1,x_0,y_0\right)}{1-g^2 g^{2}_0 (\omega^{n}_{p,k})^2}\right]\\
& = \Theta (p - k) \frac{2 \pi}{p} + \Theta (k - p) \frac{2 \pi(g g_0)^2 k}{\left[(g g_0)^2 k^2 - 1\right]}.
\end{split}
\end{equation}
Therefore, the mass function reads
\begin{widetext}
\begin{equation}
\begin{split}
&\Sigma(p)=\frac{C_1 T}{(2\pi)^2} \int_0^{\infty}\!\!\!\! d k \,\,\,\frac{k\,\,\Sigma(k)}{\pi^2 T^2+k^2+\Sigma^2(k)}\left[ \Theta (p - k) \frac{2\pi}{p} + \Theta (k - p) \frac{2 \pi}{k}\right]\\
& + \frac{C_2 T}{(2\pi)^2} \! \int_0^{\infty}\!\!\!\! d k\,\,\,  \frac{k\,\,\Sigma(k)}{ \pi^2 T^2+k^2+\Sigma^2(k)}\left[\Theta (p - k) \frac{2 \pi}{p} + \Theta (k - p) \frac{2 \pi(g g_0)^2 k}{\left[(g g_0)^2 k^2 - 1\right]}\right],
\label{masselliptic}
\end{split}
\end{equation}
\end{widetext}
which is Eq.~(\ref{auxST}).

\end{document}